\documentclass[12pt,preprint]{aastex}
\def\aca{Acta Astron.}

\shorttitle{Fold Caustic Diffraction Effects} \shortauthors{Zabel and Peterson}

\begin{document}

\title{Extended Source Diffraction Effects Near Gravitational Lens Fold Caustics}

\author{S.A. Zabel\altaffilmark{1} and J.B. Peterson\altaffilmark{2}}
\affil{Physics Department, Carnegie Mellon University, Pittsburgh, PA 15213}
\altaffiltext{1}{zabel@cmu.edu}
\altaffiltext{2}{jbp@cmu.edu}
\begin{abstract}
Calculations are presented detailing the gravitational lens diffraction due to the steep brightness gradient of the limb of a stellar source. The lensing case studied is the fold caustic crossing. The limb diffraction signal greatly exceeds that due to the disk as a whole and should be detectable for white dwarf sources in our Galaxy and its satellites with existing telescopes. Detection of this diffraction signal would provide an additional mathematical constraint, reducing the degeneracy among models of the lensing geometry. The diffraction pattern provides pico-arcsecond resolution of the limb profile.
\end{abstract}


\section{Introduction}
In gravitational lensing two new images appear when a source crosses into a caustic curve. The two new images initially appear together on the critical curve with the same path length from source to observer. As the source moves across the caustic the images separate and their respective path lengths change. In microlensing multiple images of a source can not be spatially resolved so their amplitudes are combined at the detector producing diffraction and interference. The subject of this paper is the diffraction pattern that appears near the fold caustic crossing that occasionally occurs during microlensing events. This pattern is proportional to an Airy function, and the Airy diffraction length $d_f$ determines the periodicity of the intensity oscillations. For sources whose size is less than this characteristic length diffraction effects appear at full intensity. For sources larger than this length diffraction becomes diminished. In this paper we show that for stellar mass main sequence and white dwarf sources significant diffraction fringes will be present in the observed intensity. We show detectability of a diffraction signal is dependant on the limb profile rather than source size. We apply the results to two observed caustic crossing events, the Galactic bulge event 96-BLG-3 and the Small Magellanic Cloud (SMC) event 98-SMC-1 using both the observed sources and hypothetical white dwarf sources.

We begin with the point source approximation, describing the lensing geometry and the lens equation. Next we describe the time delay surface near a fold type caustic. Using this time delay surface we calculate the diffraction integral. We then integrate over extended source profiles. We apply these results to various astrophysical situations showing examples where the limb effect is significant and we discuss examples where it is not. We compare two types of source models, a gaussian brightness profile and a brightness profile that mimics the sun.

Diffraction and interference effects in gravitational lensing have been considered by several authors. The theory of gravitational lensing was first derived in a simple manner by \citet{ref64}. Some effects of caustic crossings are discussed in \citet{oha82}, \citet{bla86},  and \citet{goo87}. Wave effects and diffraction in gravitational lensing are discussed in \citet{deg86} and \citet{daw86}. Interference is discussed in \citet{pet91} and \citet{bor97}. Diffraction effects in the frequency domain are discussed in \citet{gou92}, and \citet{ulm95} also includes a discussion on white dwarf effects in the Large Magellanic Cloud. Lensing near critical points is discussed in \citet{sta93}. Diffraction near fold caustics for gaussian sources is discussed in \citet{jar95}. Determination of source limb properties from microlensing is discussed in \citet{gou01}.
\section{Lensing Geometry}
The gravitational thin-lens geometry for a point source and a point mass lens is shown in Figure 1. Although the light is actually in gravitational interaction with the lens over the entire path, in the thin-lens approximation there is a single scattering of the light. A ray of light moves from the source at point S to the image (or lens) plane. In the plane of the lens the ray interacts gravitationally with the lens mass at impact parameter $\xi$. The ray is bent by an angle $\alpha$ and travels to the observer at point O. The source position (in the absence of a lensing mass) is at an angle $\beta$ from the line of sight to the lens. The observer sees the ray at an angle $\theta$ to the line of sight to the lens. All of the angles in the figure are small and we assume a Euclidean metric. The distance of the lens from the observer is $D_d$, the distance of the source from the observer is $D_s$, and the distance of the lens from the source is $D_{ds}$.

General relativity predicts that a light ray passing a spherical body of mass $M$ at a minimum distance $\xi$ is deflected by the "Einstein angle"  \begin{equation}\alpha=\frac{4GM}{c^2\xi}=\frac{2R_s}{\xi}\end{equation} provided the impact parameter $\xi$ is much larger than the corresponding Schwarzschild radius \begin{equation}R_s=\frac{2GM}{c^2}\end{equation} In the special case of perfect alignment, $\beta=0$ so the source, lens, and observer positions all occur on a line. By symmetry there will then be a ring of images in the lens plane at an angle $\theta$ to the line containing the lens and source. This "Einstein ring" will be at an impact parameter \begin{equation}R_E=\sqrt{2R_s\frac{D_dD_{ds}}{D_s}}=\sqrt{2R_sD}\end{equation} from the lens using the definition $D \equiv ( D_d D_{ds} ) / D_s$. $R_E$ is called the "Einstein radius".

For a point source the ray geometry obeys the lens equation\begin{equation}\beta D_s = \frac{D_s}{D_d}\xi-\frac{2R_S}{\xi}D_{ds}\end{equation}Another path, not shown in Figure 1, passes on the other side of the lensing mass. This path is described by a second solution to the lens equation.

In microlensing surveys, lensing light curves are recorded and then an attempt is made to determine the mass, position, and relative transverse velocity of the lens. In order to fully model a microlensing event one must know all but one of the parameters in equation (4). In many cases the lens is not directly observed so the lens position is not known. This leads to degenerate solutions of the lens equation and incomplete information on lensing events. The lens equation is discussed in detail in \citet{sef92} and \citet{pet01}.
\section{Fold Caustics}
If the lens is not a point-like object but consists of several point like objects (i.e. a star with planets) the symmetry is usually broken and there can not be a perfect "Einstein ring". Instead there can be multiple paths from the source, through the image plane, to the observer, each producing an image. If the lens system and source are sufficiently misaligned there will only be the image of Figure 1. But as the source and lens move close to alignment new images can appear. For a fixed observer and lens, the set of points in the source plane where the number of images seen by the observer changes forms a closed curve called a caustic. The simplest and most common type of caustic is the fold caustic. This is the type of caustic crossing we consider in this paper. The number of images always changes by two when a source crosses a fold caustic and the newly created images are generally highly magnified. The newly created images are distinct from the images that existed before the source crossed the caustic. 

The set of points in the image plane where the corresponding images are created or destroyed is called a critical curve. Whenever two new images are created they appear together on the critical curve. The images initially have the same path length from source to observer and therefore interfere constructively. But, as the source moves across the caustic, the images separate. If the images are unresolved, as in microlensing, the combined image intensity oscillates as a diffraction pattern. See \citet{boz00} for a description of the trajectories of the images in caustic crossing events.

It is standard to choose a coordinate system where the fold caustic lies along the $y_s$ coordinate, the $x_s$ coordinate is orthogonal to the caustic, and $x$ and $y$ describe the image coordinates. Near a fold caustic the non-cosmological time delay surface can be approximated locally by \citep{jar95} \begin{equation}\tau=\frac{1}{Dc}\left(\frac{1}{48d_c}x^3 + y^2 - x_sx-y_sy+\frac{1}{2}x_s^2+\frac{1}{2}y_s^2\right)\end{equation} The quantity \begin{equation}d_c=\alpha\ R_E\end{equation} is the Einstein radius of the lens modified by the factor \begin{equation}\alpha\equiv\langle K^2\rangle / \langle m\rangle\end{equation} Here $K$ is called the flux factor of the source and $\langle m\rangle$ is the averaged mass of the lensing objects in solar units. The value of $\alpha$ is $\sim$ 1 and for simplicity we adopt that value. More precise calculations of alpha requires detailed knowledge of the lensing objects. See \citet{kay89}, \citet{wit90}, and \citet{wit93} for details. Equation (5) describes the time delay of an image point $(x,y)$ for a given source position ($x_s$,$y_s$). Choice of the origin of the image plane is arbitrary for the purposes of this paper since diffraction calculations involve integration over the entire image plane.

In the geometrical optics limit, the magnification of a point source just inside a caustic varies as \begin{equation}A=A_{min}+(d_c/x_s)^\frac{1}{2}\end{equation} where $x_s$ is the distance of the source from the caustic. On the outside of the caustic there is only the original image and $A=A_{min}$. For a point source on the caustic the magnification is formally infinite. However, as seen below, using an extended source and applying diffraction theory, the magnification remains finite.
\section{Diffraction}
In scalar diffraction theory the magnification of a set of images can be written as the modulus squared of the amplitude \begin{equation}\Psi(x_s,y_s)=\frac{-i}{2\pi d_F^2}\int^{+\infty}_{-\infty}dx\int^{+\infty}_{-\infty}dy \ exp (i\omega\tau)\end{equation} Here the Fresnel length is \begin{equation}d_F= \left(\frac{\lambda D}{2\pi}\right)^\frac{1}{2} \end{equation} The constant in front of the integral normalizes the magnification to one in the absence of a lens. For a point source near a fold caustic we use the time delay surface given in equation (5) in the diffraction integral in equation (9). This approximation only includes the contribution of the images that were created in the caustic crossing and so is only valid near the caustic crossing event itself. We also adopt a value of $y_s=0$ for the distance of the source from the caustic in the direction parallel to the caustic. If the size of the source is small compared to the radius of curvature of the caustic curve this is a valid assumption. 

Evaluation of the integral leads to \begin{equation}\left| \Psi(x_s,y_s) \right| = 2^\frac{4}{3}\pi^\frac{1}{2}\left( \frac{d_c}{d_F} \right)^\frac{1}{3} \left| Ai \left( \frac{-x_s}{d_f} \right) \right|\end{equation} where $Ai$ is the Airy function and \begin{equation}d_f=\frac{d_F}{2^\frac{4}{3}} \left(\frac{d_F}{d_c}\right)^\frac{1}{3}\end{equation} is the characteristic scale for the argument change, or the Airy diffraction length. Equation (11) is a factor of two larger than the result obtained by \citet{jar95} because we have corrected a minor error in the limits of the integral. This increases the magnification by four.
\section{Extended Source Effects}
The physical optics calculation for an extended source requires integration over both the source and image planes. Assuming incoherent emission, the contribution to the integrated flux of an extended source from the new images near a caustic crossing is given by \begin{equation}F=\int^{+\infty}_{-\infty}dx_s\int^{+\infty}_{-\infty}dy_s I\left(x_s,y_s\right)\left|\Psi\left(x_s,y_s\right)\right|^2\end{equation} The integral is over the source plane, and each point in the source plane contributes based on an integral in the image plane from equation (9). \begin{math}I(x_s,y_s)\end{math} is the surface brightness profile of the source. To proceed further we must choose a model of the surface brightness of the source.

Because the limb brightness gradient will be the dominant factor controlling the amount of diffraction, we need a source model with a realistic limb profile. There is little information on limb gradients for any star except the sun and a few nearby giant stars. A reasonable model radial profile of the sun can be seen in Figure 2. This profile of the sun is that described in \citet{rhi02} and \citet{car80}. The intensity is normalized to unity at the center. Moving out from the center the intensity falls off as a cosine. There is an abrupt jump to 40$\%$ of the central intensity at the stellar radius.

Assuming spherical symmetry the surface brightness of any point on the surface of the sun is then described by
\begin{equation}
\frac{I(\theta)}{I_0} =\cases{0.4+0.6\ cos(\theta) & 
\qquad$r\leq r_{\odot}$ \cr
0 & \qquad$r > r_{\odot}$ \cr}
\end{equation}
The angle $\theta$ is related to the distance $r$ from the center of the star as seen on the projected solar disk by\begin{equation}\theta(r)=sin^{-1}\frac{r}{r_{\odot}}\end{equation}

This model is a good approximation to the observed solar profile. We can use the discontinuous approximation because for all cases in this paper the width of the region over which the limb intensity increases is much less than the diffraction length $d_f$ \citep{lit83}, \citep{kun98}.

When considering white dwarf sources we scale this solar profile to the smaller radius of the white dwarf. This assumes that the strong gravitational field at the white dwarf surface maintains a steep density gradient in the white dwarf atmosphere.

There are two scales for diffraction effects when a solar-profile source crosses a caustic. The first is the disk diffraction effect of the entire source as it moves inside the caustic region. Just inside the caustic, when the source radius is much smaller than $d_f$, all components of the source are localized within one fringe of the Airy function. As the source moves further into the caustic the entire source contributes nearly identical values of the Airy function and there is full amplitude diffraction. When a source is larger than $d_f$ different regions of the source will be at different phases within the Airy function. This leads to an averaging, so for large sources diffraction effects become diminished.

The second effect is a limb effect. As the source first begins to cross the caustic only its limb is magnified. This is shown in Figure 3. Most of the source has not crossed the caustic and those points are not yet additionally imaged. So when a source of any size first crosses a caustic its limb will always show diffraction effects. However, even for the limb of the source, the averaging effect will begin to take place as the limb moves into the caustic. The diffraction oscillations will begin to average out when the limb has passed many characteristic lengths $d_f$ into the caustic.

In order to extract the small limb diffraction signal from the large unlensed contribution a differential observation is appropriate. One method is to rapidly measure differences of intensity as the source moves across the caustic. By sampling the event repeatedly such that for every time step the source has moved a distance $x_s < d_f$, and then subtracting successive intensity measurements, diffraction due to a portion of the source radius less than $d_f$ can be isolated. We have calculated this differential magnification for several cases described in the next section.
\section{Results}
To begin, we compare our calculations to previous results. Fold caustic diffraction effects for gaussian sources in the quasar lensing system Q2237+0305 have been studied by \citet{jar95}. In this event, a source at redshift $z_s$ = 1.695 is assumed to be lensed by solar mass objects at $z_d$ = 0.0394. The system has $\alpha = 0.8$. At wavelength $\lambda = 500 nm$, $d_f$ = 4.0$\times 10^8 cm$. Using these values and a gaussian radial profile (scaled to the appropriate source size), we evaluate the integral in equation (13) and obtain the results in the left panel of Figure 4. The magnifications we calculate are larger than those previously calculated due to the difference in the evaluation of the diffraction integral (equation [11]) and proper normalization of the gaussian (Jaroszy\'nski, M., and Paczy\'nski, B. 2002, private communication). For sources with a characteristic width greater than $d_f$ diffraction effects are not visible. 

To study the effects of other source profiles on diffraction we replace the gaussian profile with a model solar profile. Although the source in the case under consideration is a QSO, which does not have a solar profile, we use this profile to show that steep brightness profile gradients enhance diffraction. The right panel in Figure 4 is for solar profile sources of various radii. For sources larger than $d_f$ diffraction fringes have a greater amplitude for a solar profile source than a gaussian profile, particularly at large $x_s$. For sources with radii of 10$d_f$, for either profile, the diffraction effects are not visible. For these larger sources we need to look at the differential magnification to see fringes.

Limb diffraction is minimal for gaussian profile sources but is much stronger for a solar profile source. Figure 5 shows the magnification and differential magnification of a gaussian source with a characteristic width $\sigma$ = 10$d_f$ as it passes into a fold caustic. There are no apparent diffraction effects in Figure 5. Figure 6 shows the magnification and differential magnification of a solar profile source with a much larger radius (1000$d_f$) as it crosses a fold caustic. Figure 6 shows diffraction effects in the differential magnification plot. For sources 1000 times the Airy diffraction length, even though the whole-disk diffraction effect is almost completely averaged out, a substantial limb effect can be seen. The steep gradient of the scaled solar profile produces the diffraction signal seen in Figure 6. Because a gaussian profile has much smaller gradients, or alternatively because the width of the limb region is larger than $d_f$, the limb effects are severely diminished. 

There are several ongoing microlensing surveys using sources in our Galaxy and its satellites. These surveys occasionally report caustic crossing events. The MACHO/GMAN collaboration \citep{alc00}, the PLANET collaboration \citep{alb01}, the OGLE experiment \citep{uda92,uda97}, and the MPS collaboration \citep{rhi98} are all currently using sources in either the Galactic bulge or Magellanic clouds. We now adopt lensing parameters applicable to these surveys.

The event 96-BLG-3 was the third bulge caustic crossing event detected by the MACHO project in 1996. A $1.1R_{\odot}$ G0 star at $8.5kpc$ was lensed by a $1.61M_{\odot}$ binary lens at $7.1kpc$. The velocity of the source perpendicular to the caustic was $103km s^{-1}$. This event is described in \citet{alc00} and \citet{len96}. The event 98-SMC-1 was the first SMC caustic crossing event detected by the MACHO project in 1998.  The time of the second caustic crossing (passing out of the multiple image region) was predicted far enough ahead of time that all of the groups listed above were able to observe this event. A $1.1R_{\odot}$ star at $62.5kpc$ was lensed by a $0.36M_{\odot}$ binary lens at $58kpc$. The velocity of the source perpendicular to the caustic was $76km s^{-1}$. This event is described in \citet{alc00} and \citet{rhi99}. We use these two observed events as our canonical examples.

First we take the 96-BLG-3 case. Evaluating the diffraction integral we get the results in Figure 7. Since $d_f$ is the typical length for diffraction effects this is the resolution obtained by measuring these effects. For this system $d_f = 9.6\times10^5$cm, or 7.5 pico-arcsecond on the sky. For the quoted source velocity of $103km s^{-1}$ the initial period of oscillation is $1/10s$. The disk magnification reaches a maximum of $\sim 78$. Although this is slightly less than the observed value of $\sim 120$ \citep{alc00} the source passed near a cusp and therefore the peak magnification would be slightly higher than we predict when assuming a fold crossing. As seen in the differential magnification plot, the amplitude of the diffraction fringes is $\sim 5\times10^{-5}$ of the unlensed source intensity.

Next we take the 98-SMC-1 case. Evaluating the diffraction integral we get the results in Figure 8. This system has $d_f = 2.3\times 10^6 cm$ and translates to 2.49 pico-arcsecond on the sky. For the quoted source velocity of $76 km s^{-1}$ the initial period of oscillation is $1/3 s$. The disk magnification reaches a maximum of $\sim 78$. This is in good agreement with the best fit to the data of $\sim70$ \citep{rhi99}. As seen in the differential magnification plot, the amplitude of the diffraction fringes is $\sim 1\times10^{-4}$ of the unlensed source intensity.

The signals produced in the 96-BLG-3 and 98-SMC-1 cases are too small to be detected with available technology. Since compact objects produce strong diffraction effects we now use white dwarfs as the sources in our models. We replace the main sequence sources in our two previous cases with a $1 M_\odot$ DA white dwarf similar to Sirius B, with a radius of $0.008 R_\odot$. Using this white dwarf as the source object in the 96-BLG-3 event we evaluate the diffraction integral and get the results in Figure 9. The data points were separated by source displacement steps of size $d_f/4$. The disk magnification reaches a maximum of $\sim 800$. In the differential plot, the amplitude of the diffraction fringes is $\sim 8\%$ of the unlensed source intensity. Substitution of a white dwarf source does not change the initial period of oscillation or $d_f$.

Finally we take the case where the same white dwarf is the source object in the 98-SMC-1 lensing case. Evaluating the diffraction integral we get the results in Figure 10. The disk magnification reaches a maximum of $\sim 1100$. As seen in the differential magnification plot, the amplitude of the diffraction fringes is $\sim 40\%$ of the unlensed source intensity.
\section{Discussion}
For the cases we have considered the whole disk diffraction signals are undetectable. For main sequence stars in the Galactic bulge and SMC the limb diffraction intensity is $\sim 10^{-4}$ the unlensed source intensity, also undetectable with current technology. However, for white dwarf sources in the Galactic bulge and SMC limb diffraction is 10 to 40$\%$ of the unlensed source intensity. It is interesting that diffraction is more readily detected in small sources than in large ones. This is opposite the usual bias in astronomical observations.

So far there have not been any reported white dwarf caustic crossing events. The reported events have much brighter source stars. However, if white dwarf events are discovered in the Galactic bulge or Magellanic clouds it should be possible to detect a diffraction signal.

We have considered only a single wavelength in this analysis. However, the effects we describe may be easier to detect if broad optical bands can be used. All the diffraction patterns presented here scale with wavelength so if a broad bandwidth is to be used it must be subdivided into spectra. The resolving power required is approximately the number of diffraction fringes from the caustic to the limb of the source. Only low resolution ($\lambda / \Delta \lambda \sim$10) is needed. However, these spectra must be recorded in rapid succession since the fringe pattern changes in a fraction of a second. 

Caustic crossing events are usually discovered after the source crosses into the caustic, which does not allow time to set up rapid spectrographic observations. Precise measurement of the event, along with modelling, allows one to predict when the source will pass back out of the caustic. The diffraction pattern produced as a source leaves a caustic is reversed in time compared to the pattern produced as it enters. The inbound crossing can be used to alert observers to record rapidly sampled spectra during the outbound crossing. 

Detection of limb diffraction would provide a measurement of $d_f$ resulting in an additional constraint which helps remove the degeneracy from the lens model. Also, accurate measurement would allow fine detail of the limb brightness profile to be studied. Diffraction patterns could provide information on distant collapsed objects with higher angular resolution than is available through any other technique.

\clearpage
\begin{figure}
\plotone{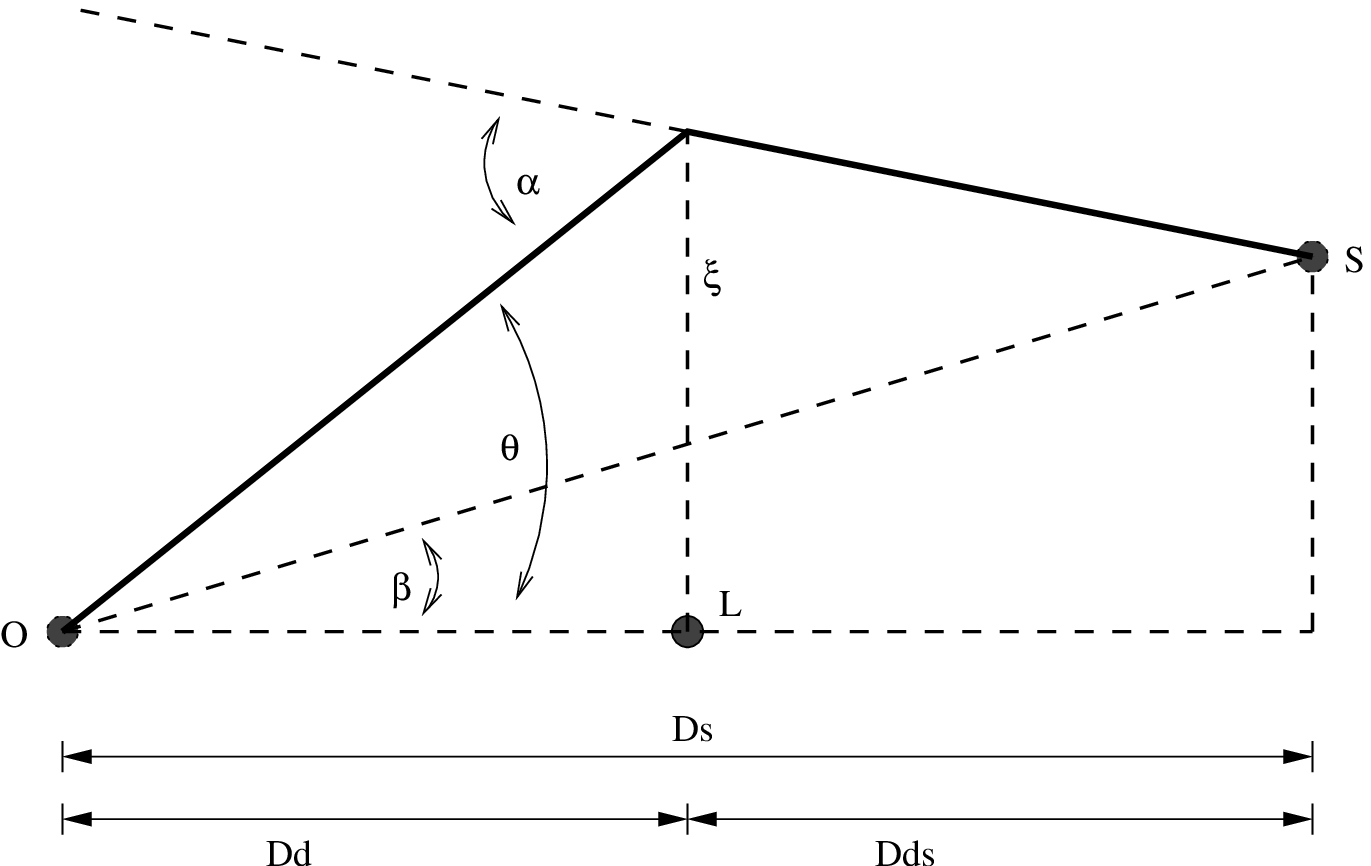} \caption{{\bfseries Point source, thin-gravitational-lens geometry.} The source is at point S, the lens at point L, and the observer at point O. The ray is bent by an angle $\alpha$ at an impact parameter $\xi$ from the lens. The observed image appears at an angle $\theta$ to the line of sight to the lens. The source position is at an angle $\beta$ to the line of sight to the lens. Another path, not shown in the figure, passes on the other side of the lens in the same plane.\label{fig1}}
\end{figure}

\begin{figure}
\plotone{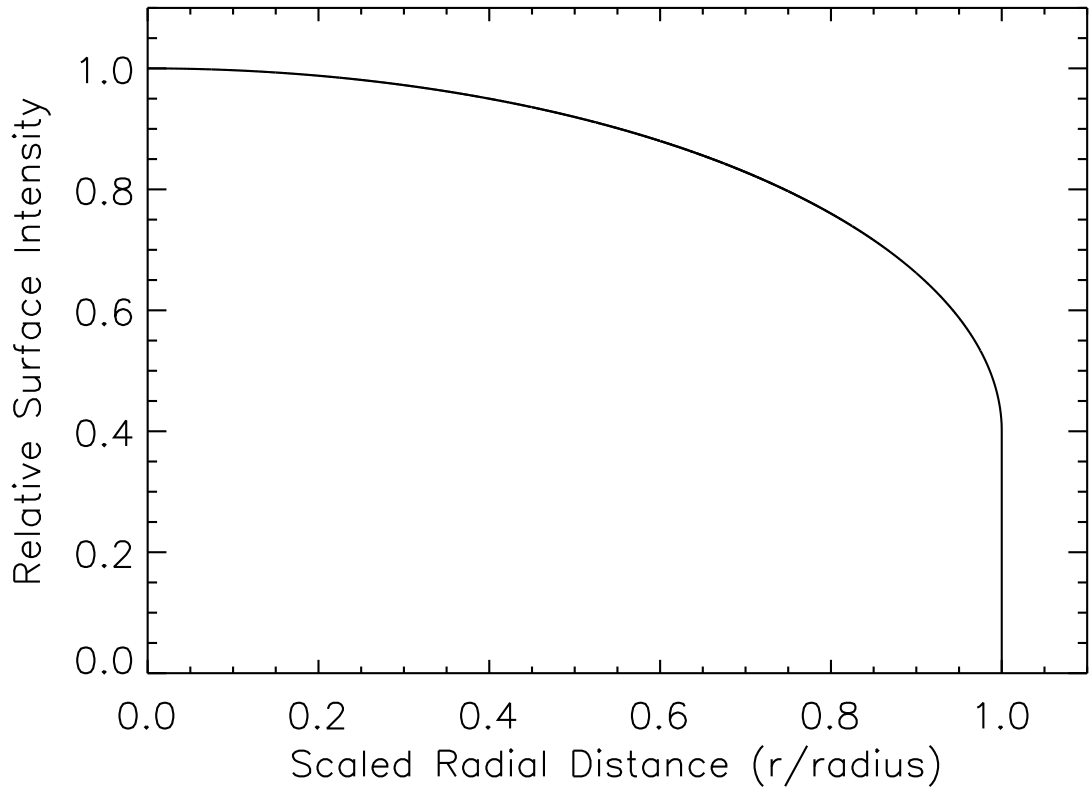} \caption{{\bfseries Model brightness profile of a solar mass main sequence star.} The intensity falls as a cosine until the limb is reached at which point it falls from 40$\%$ of the central intensity to 0.\label{fig2}}
\end{figure}

\begin{figure}
\plotone{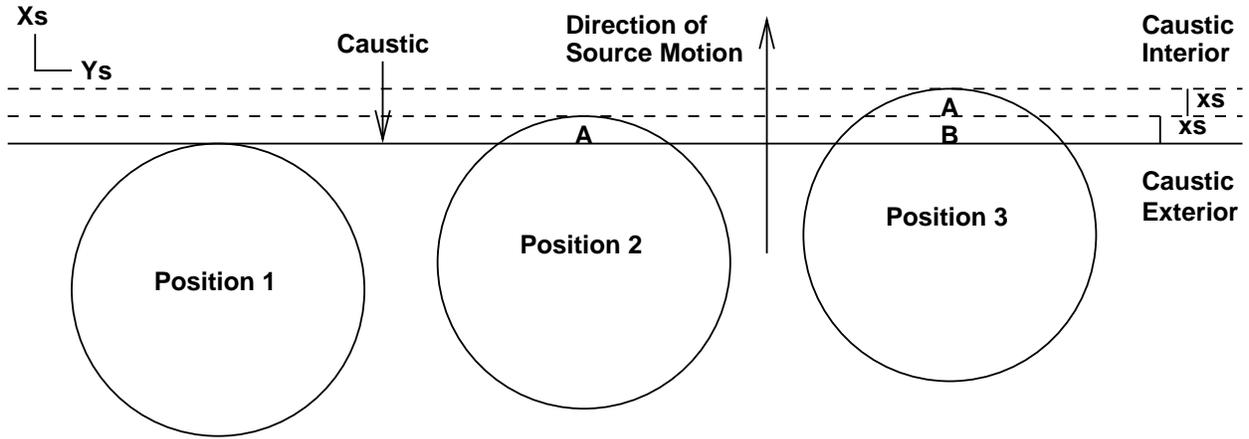} \caption{{\bfseries A circular source crossing a fold caustic.} The source coordinates $x_s$ and $y_s$ are shown. The caustic is the solid horizontal line and the source is moving up, into the interior of the caustic. The source positions are shown at three different times. The source moves a distance $x_s$ between positions. The areas A and B that move into the caustic at each step are shown. At position 2 area A produces the diffraction signal. At position 3 area B produces the dominant signal.\label{fig3}}
\end{figure}

\begin{figure}
\plottwo{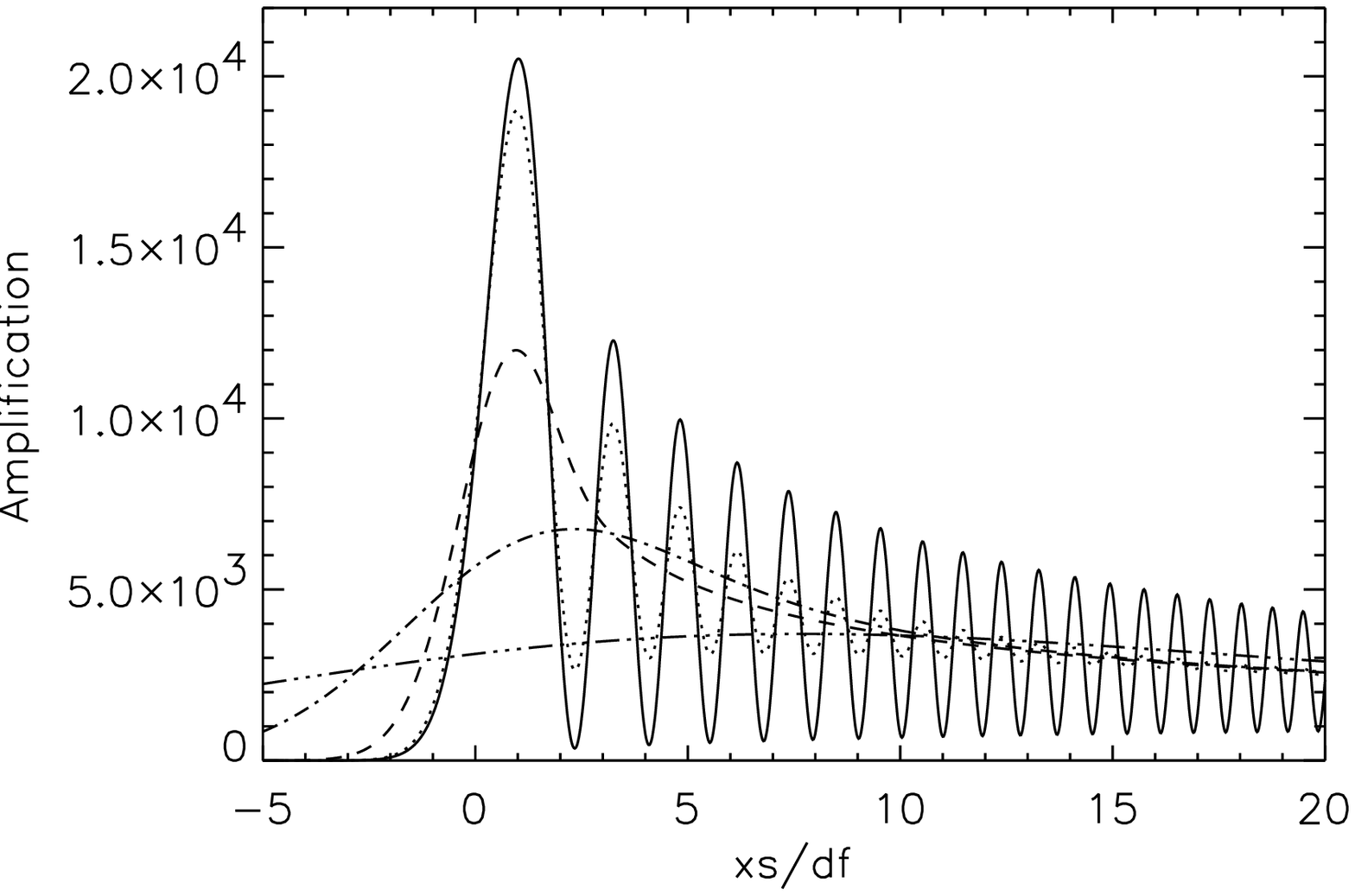}{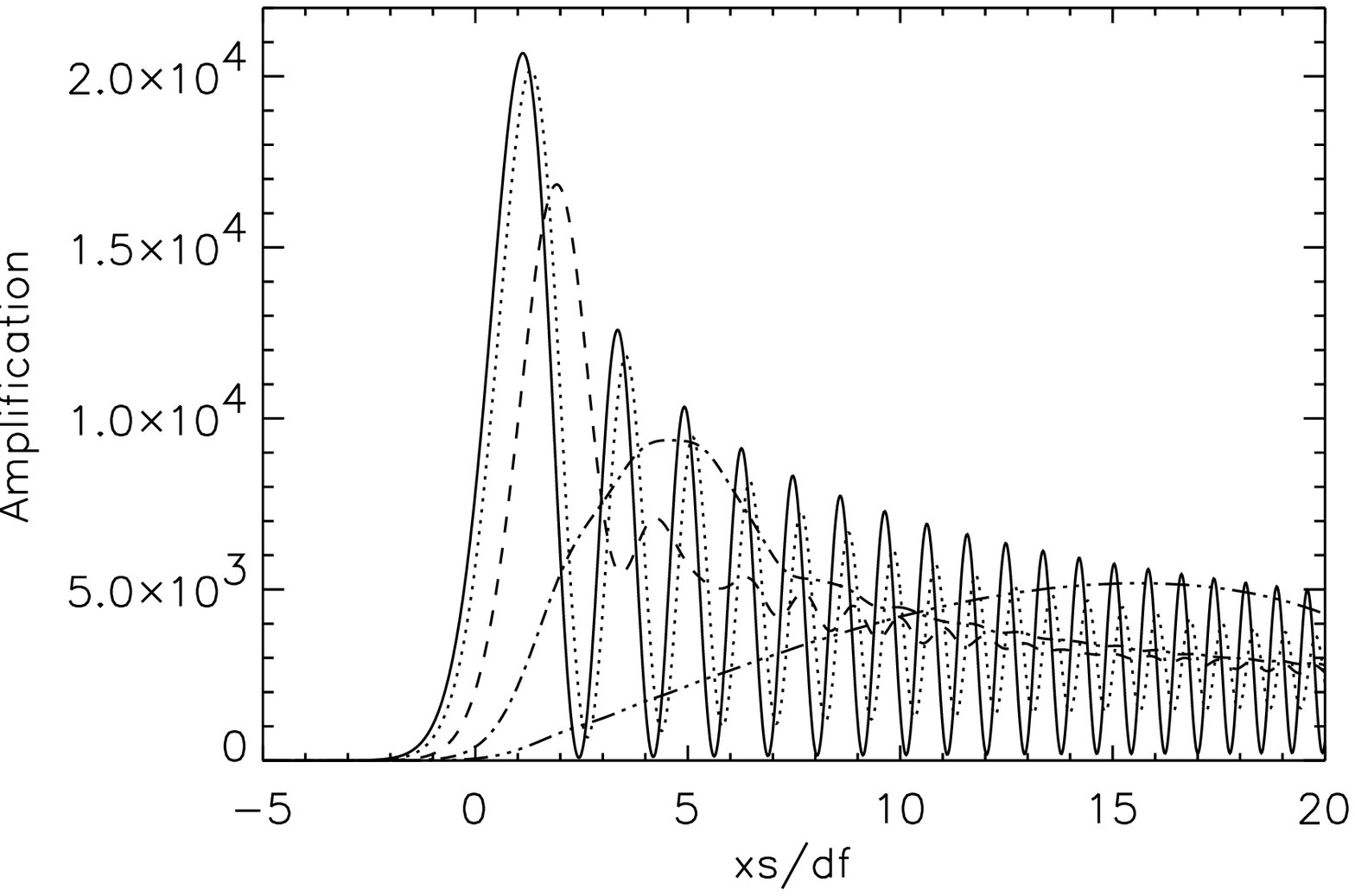} \caption{{\bfseries Magnification of various gaussian profile (left) and solar-profile (right) sources.} The magnification is plotted versus the distance the source has passed into the fold caustic. The geometry is that for Q2237+0305. The plots are for sources with characteristic widths[radii] of 0.1(solid), 0.3(dotted), 1(dashed), 3(dot dashed), and 10$d_f$ (dot dot dashed). For the solar profile, the source distance $x_s$ is measured from the limb as in Figure 3. For the gaussian profile, the source distance $x_c$ is measured from the center. In both cases the distances are scaled by the Airy diffraction length $d_f$. For a gaussian profile, for source widths comparable to $d_f$, diffraction is strongly attenuated, particularly for large $x_c$. For a solar profile source of size comparable to $d_f$ fringes are still visible at large $x_s$. In both cases, for sources with radii less than the characteristic size $d_f$, diffraction effects are visible. In both cases, for sources of size 10$d_f$, no diffraction is visible in these plots of magnification.\label{fig4}}
\end{figure}

\begin{figure}
\plottwo{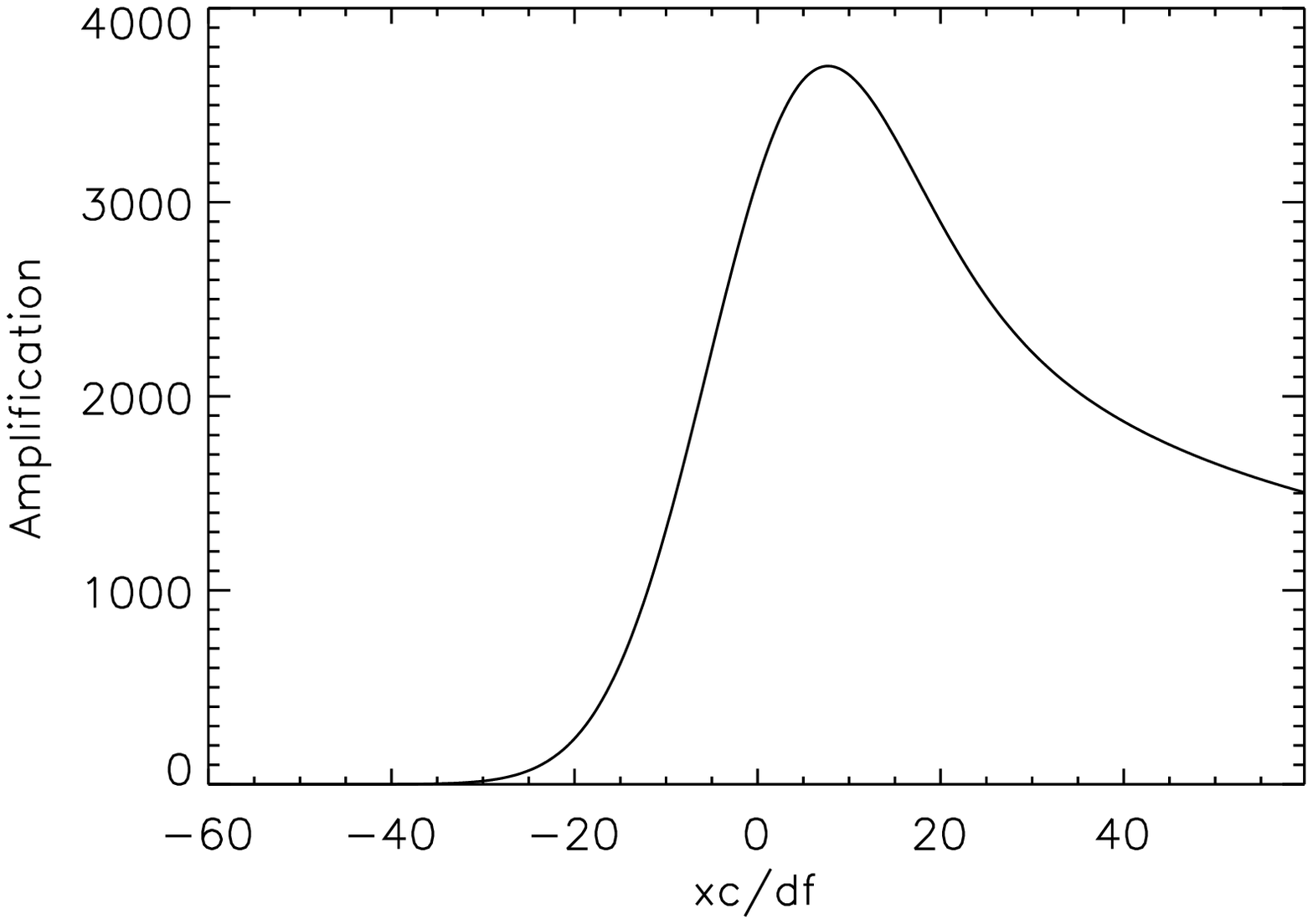}{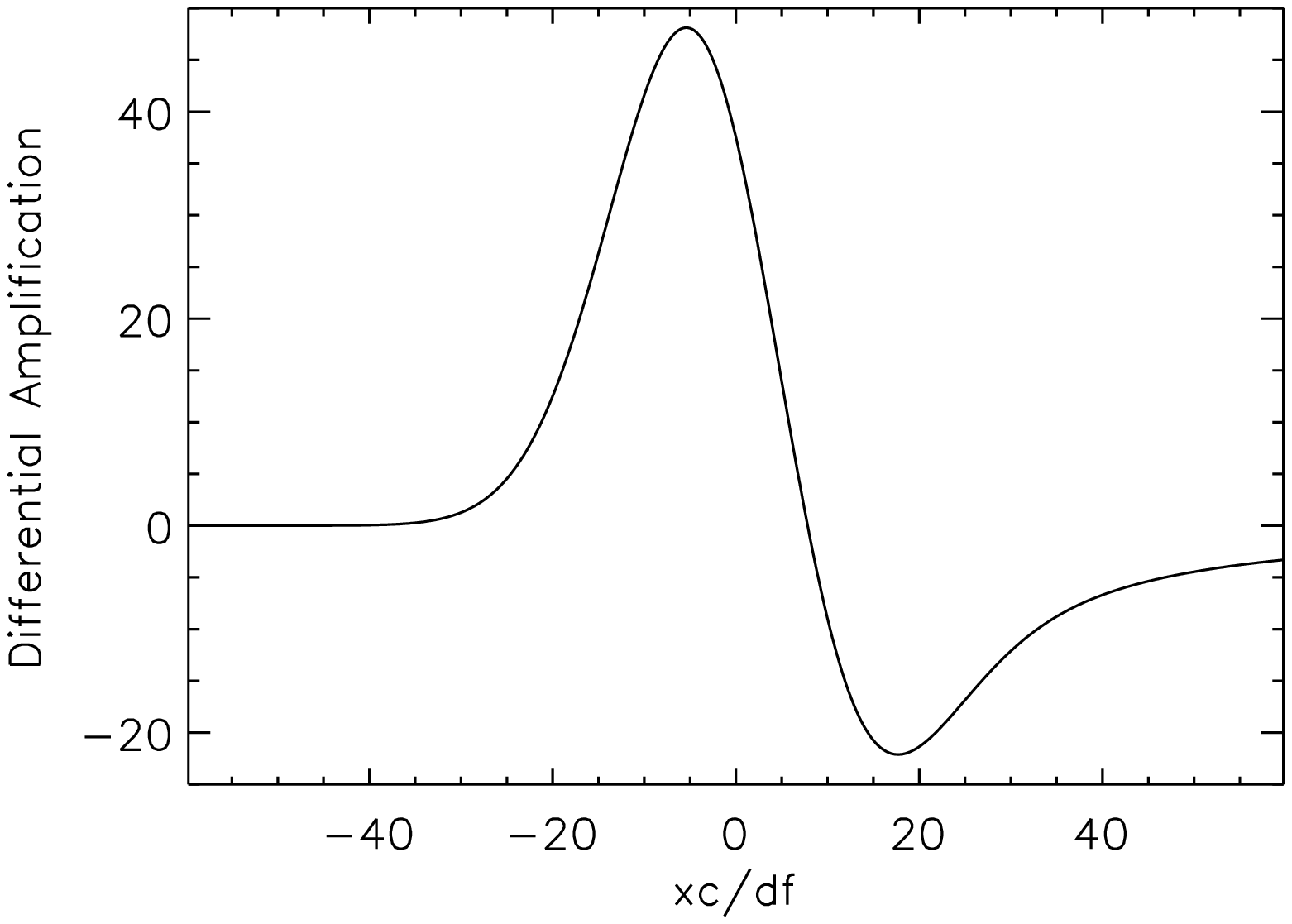} \caption{{\bfseries Magnification and differential magnification for a gaussian source.} The source width is 10$d_f$. The lensing geometry is that of Q2237+0305. The distance from the source center to the caustic $x_c$ is scaled by the Airy diffraction length $d_f$. The right panel shows differences of amplitude taken at intervals $d_f$/4. No diffraction fringes are visible.\label{fig5}}
\end{figure}

\begin{figure}
\plottwo{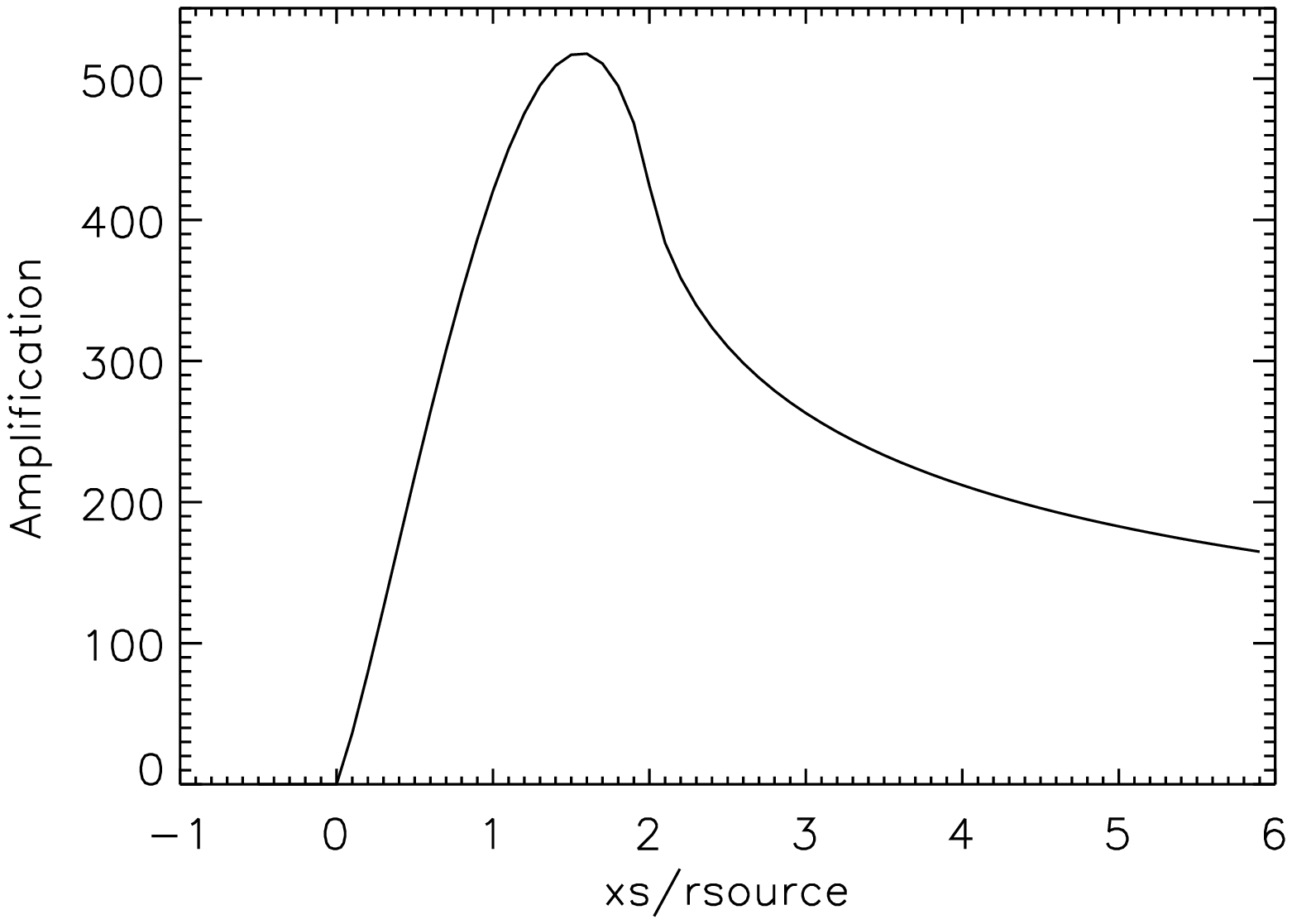}{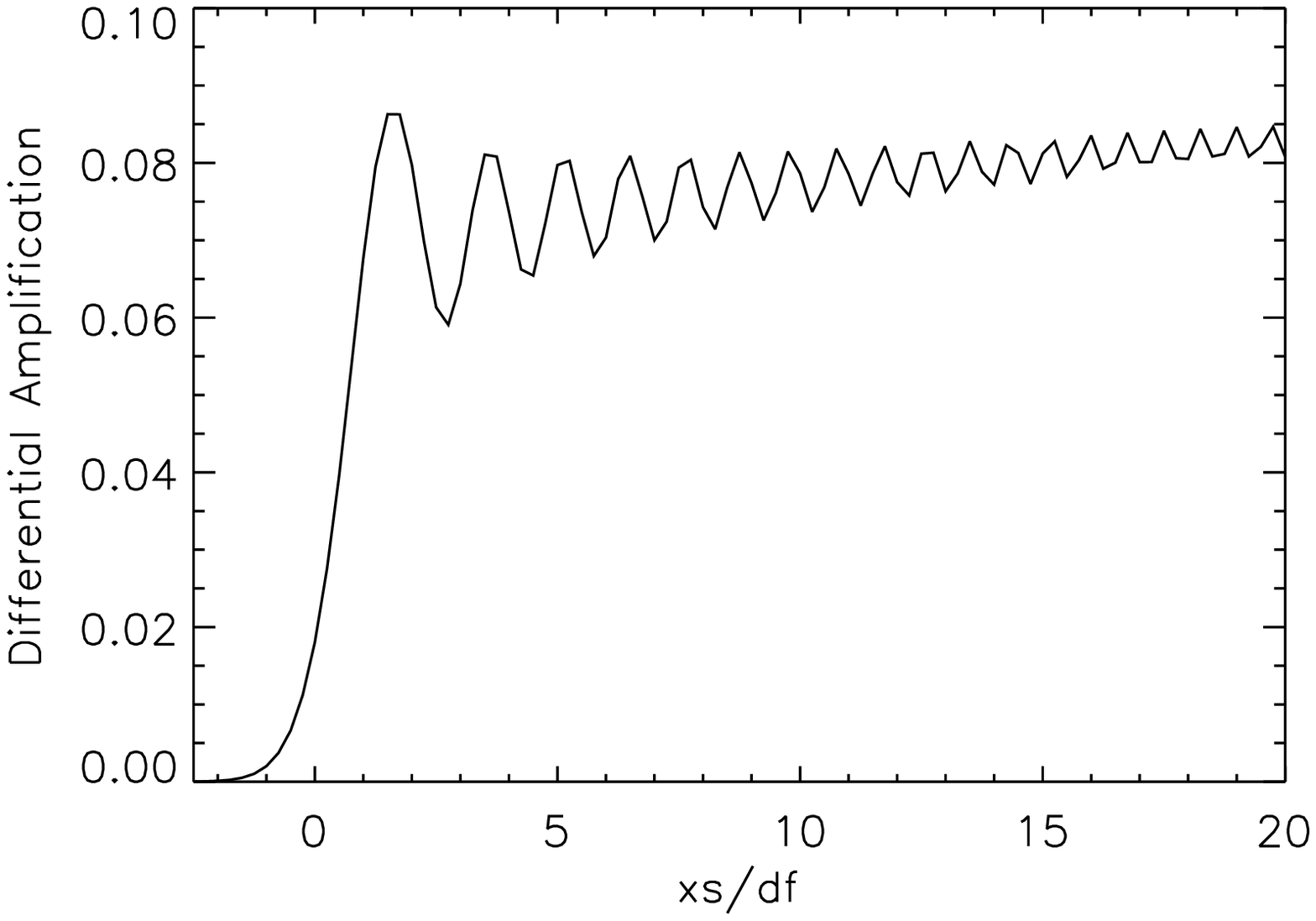} \caption{{\bfseries Magnification and differential magnification of a solar profile source.} The source has a radius of 1000$d_f$. The geometry is that of Q2237+0305. $x_s$ is the distance the limb of the source has passed into the caustic. In the magnification plot this distance is scaled by the radius, in the differential plot it is scaled by the Airy diffraction length $d_f$. The corresponding scale expansion is $r_{source}/d_f$=1000. The right panel shows differences of amplitude taken at intervals $d_f$/4. Even for a source of radius 1000 times the diffraction length, limb diffraction effects can be seen for this solar profile source.\label{fig6}}
\end{figure}

\begin{figure}
\plottwo{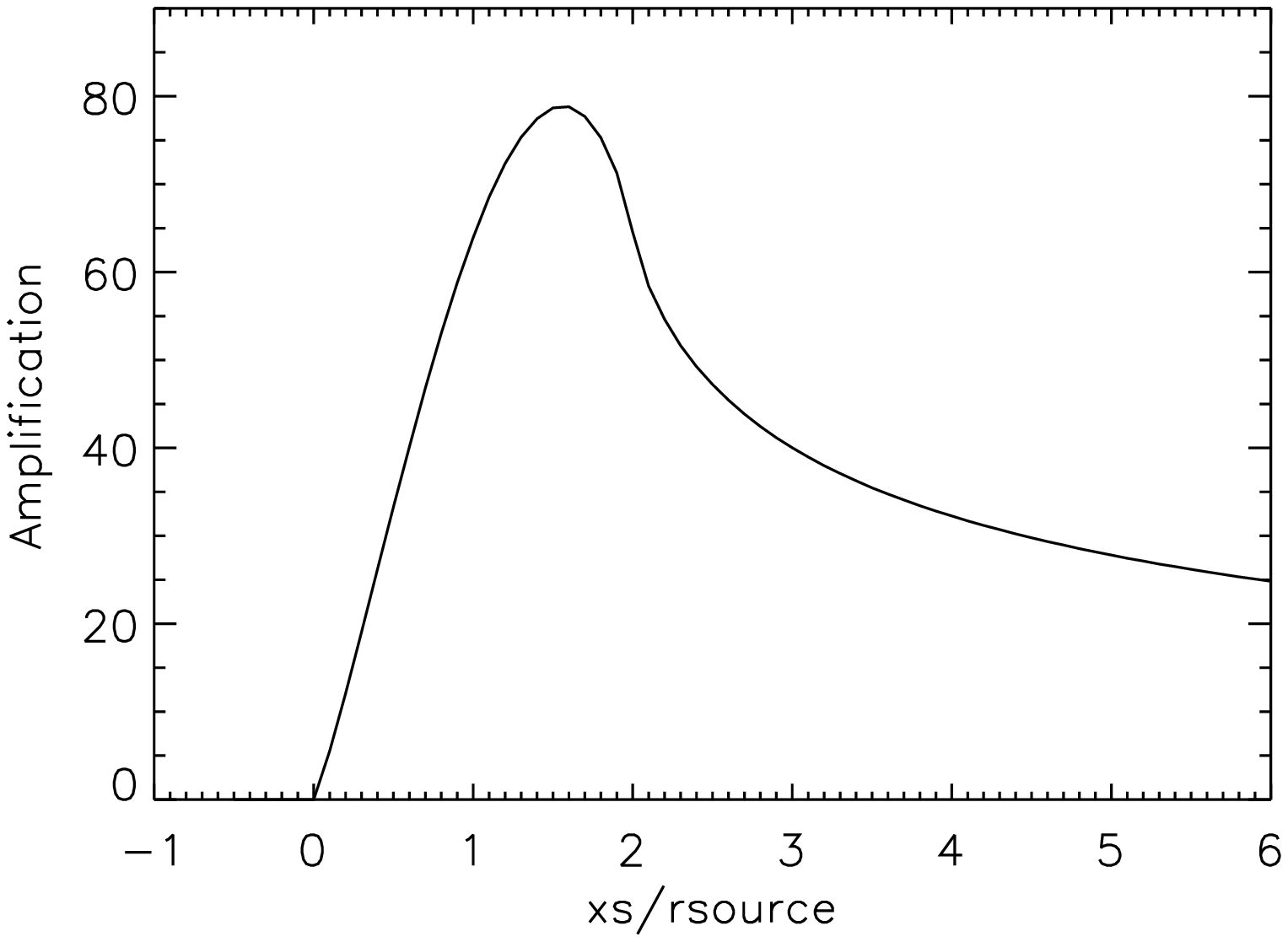}{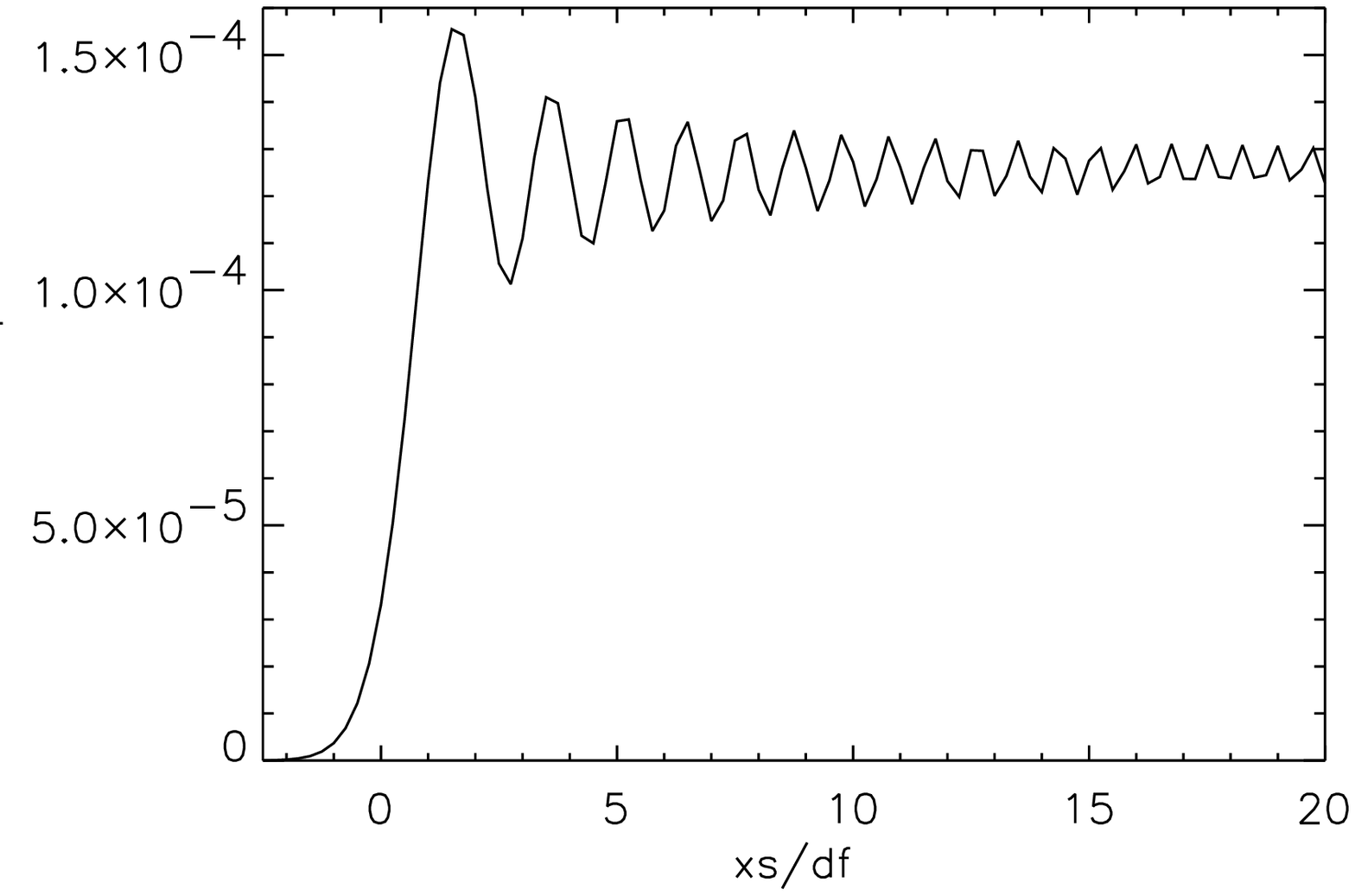} \caption{{\bfseries Magnification and differential magnification for the event 96-BLG-3.} $x_s$ is the distance the limb of the source has passed into the caustic. In the magnification plot this distance is scaled by the radius, in the differential plot it is scaled by the Airy diffraction length $d_f$. The corresponding scale expansion is $r_{source}/d_f$=$7.99\times 10^4$. The plots are sampled at $d_f$/4. The peak-to-peak amplitude of the diffraction signal in the differential plot is $\sim 5\times10^{-5}$ of the unlensed source intensity.\label{fig7}}
\end{figure}

\begin{figure}
\plottwo{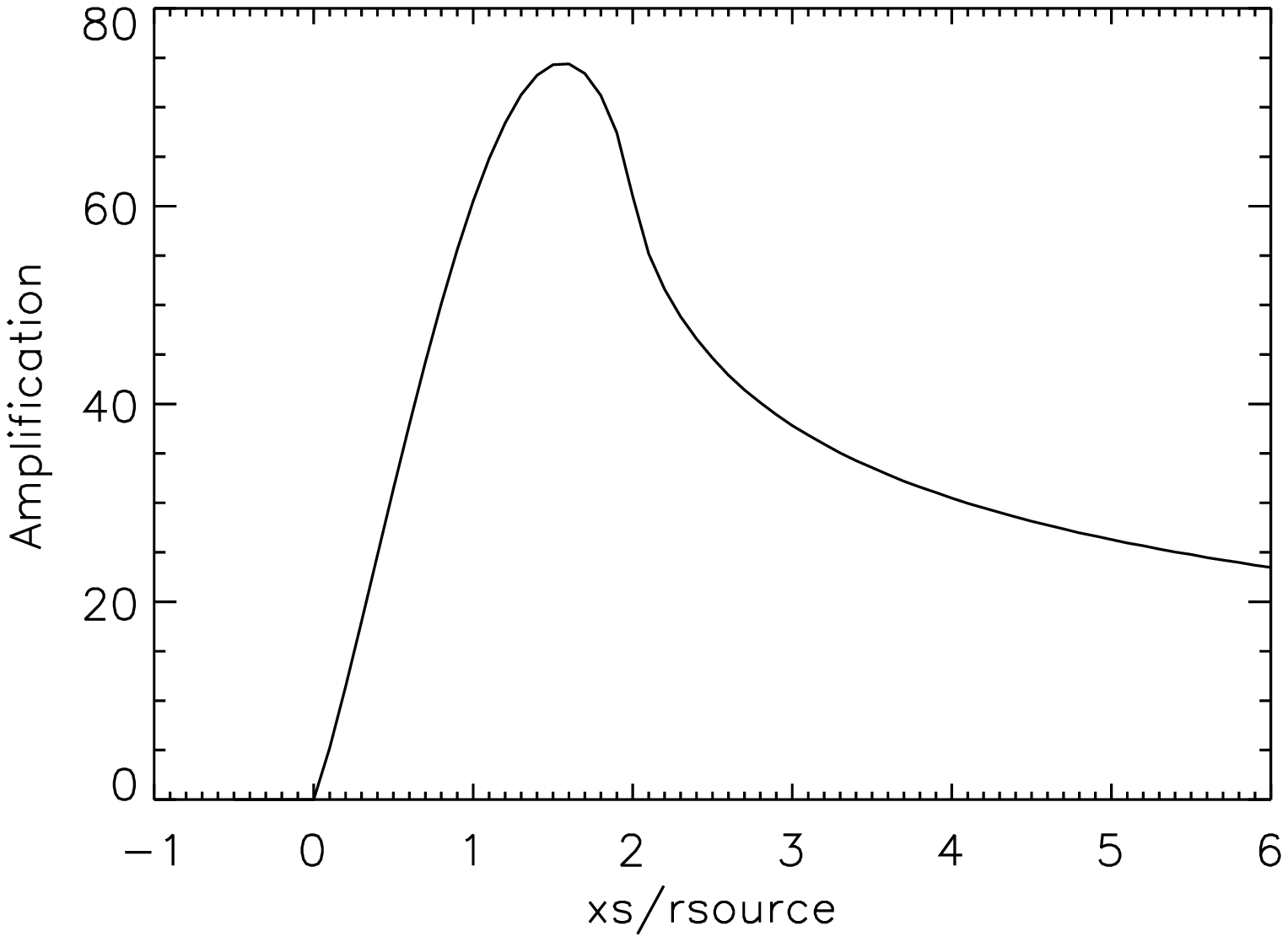}{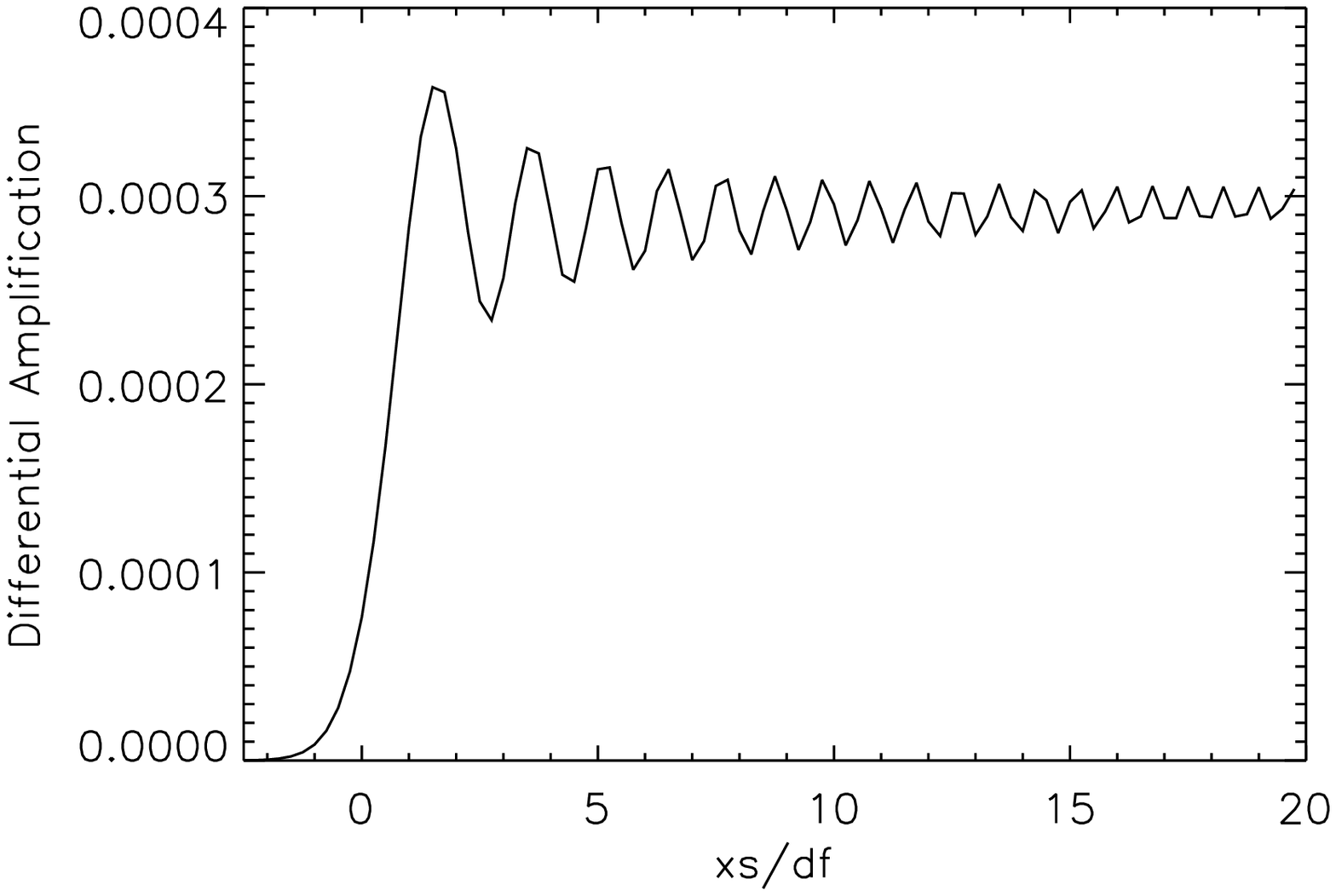} \caption{{\bfseries Magnification and differential magnification for the event 98-SMC-1.} $x_s$ is the distance the limb of the source has passed into the caustic. In the magnification plot this distance is scaled by the radius, in the differential plot it is scaled by the Airy diffraction length $d_f$. The corresponding scale expansion is $r_{source}/d_f$=$3.29\times 10^4$. The plots are sampled at $d_f$/4. The peak-to-peak amplitude of the diffraction signal in the differential plot is $\sim 1\times10^{-4}$ of the unlensed source intensity.\label{fig8}}
\end{figure} 

\begin{figure}
\plottwo{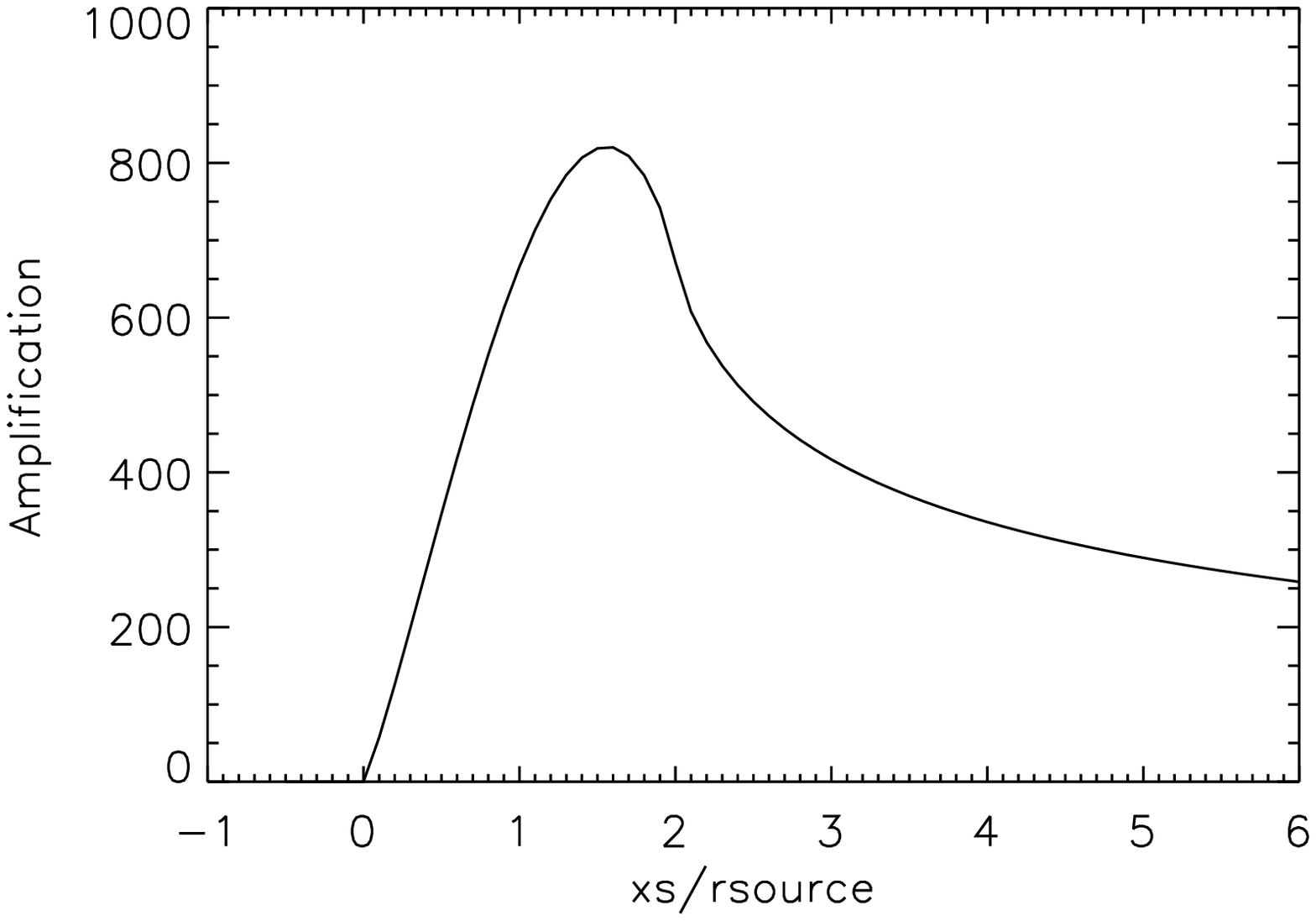}{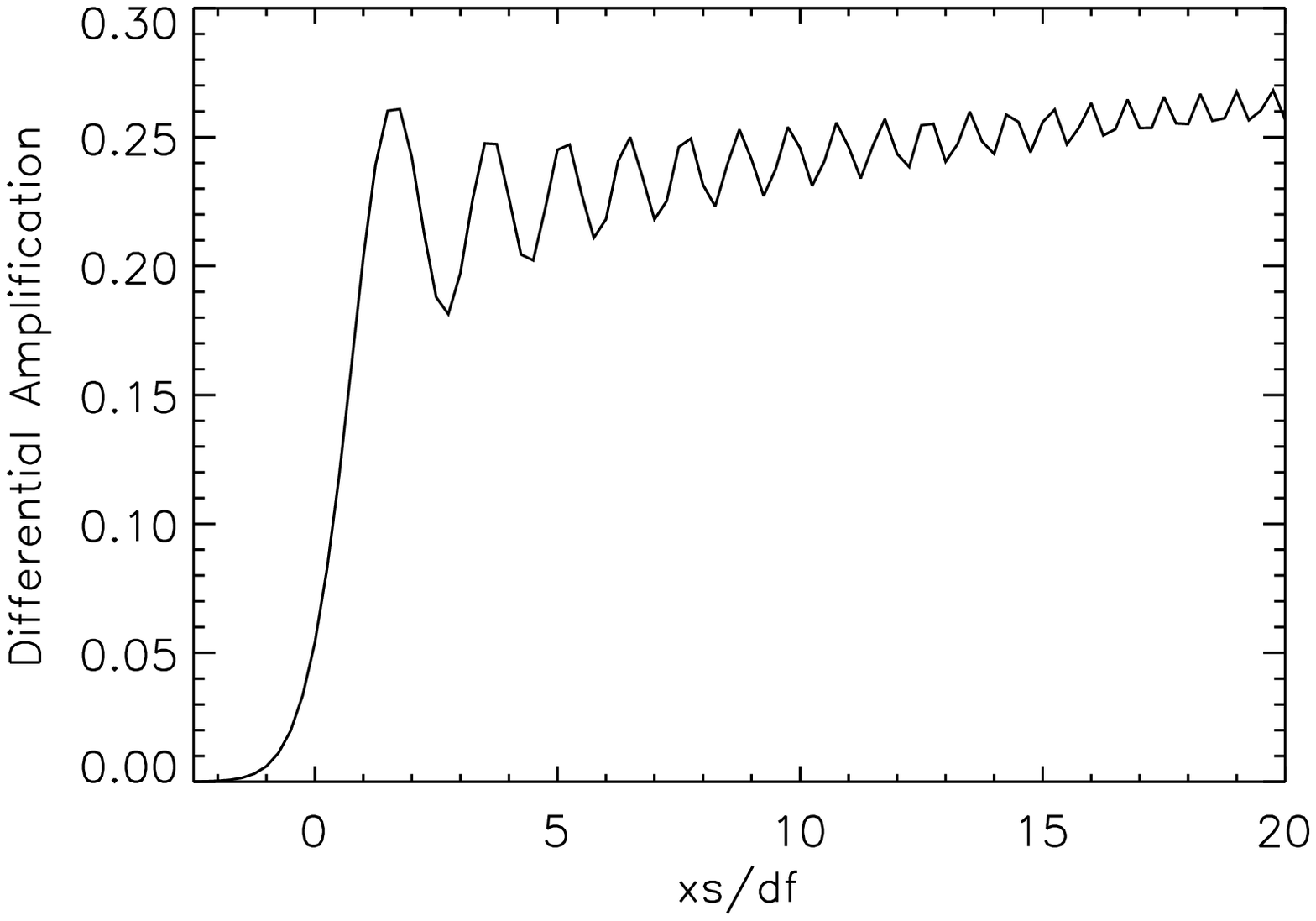} \caption{{\bfseries Magnification and differential magnification for a hypothetical white dwarf source in the 96-BLG-3 event.} $x_s$ is the distance the limb of the source has passed into the caustic. In the magnification plot this distance is scaled by the radius, in the differential plot it is scaled by the Airy diffraction length $d_f$. The corresponding scale expansion is $r_{source}/d_f$=581. The right panel shows differences of amplitude taken at intervals $d_f$/4. The peak-to-peak amplitude of the diffraction signal in the differential plot is $\sim$ 8$\%$ of the unlensed source intensity.\label{fig9}}
\end{figure}

\begin{figure}
\plottwo{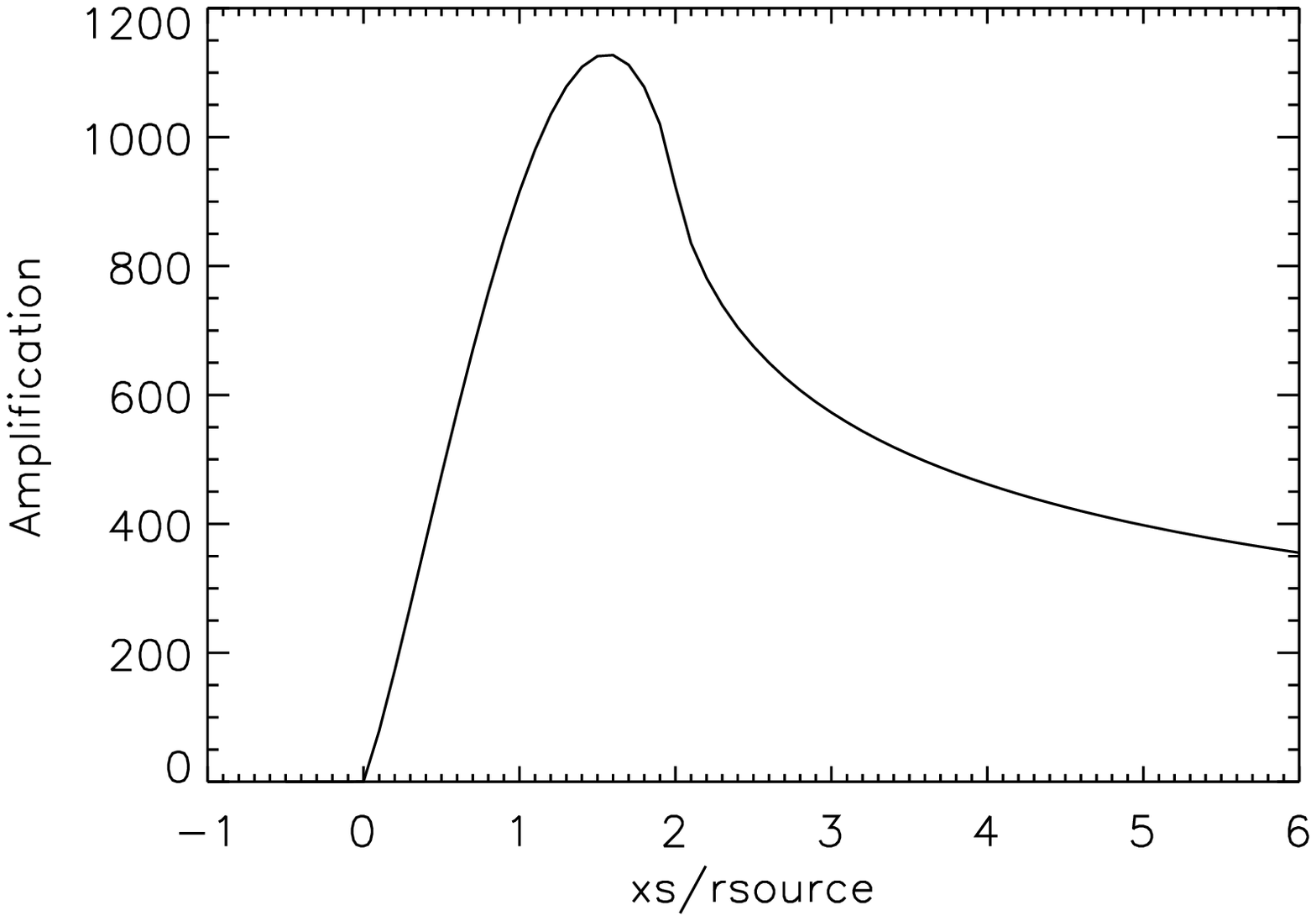}{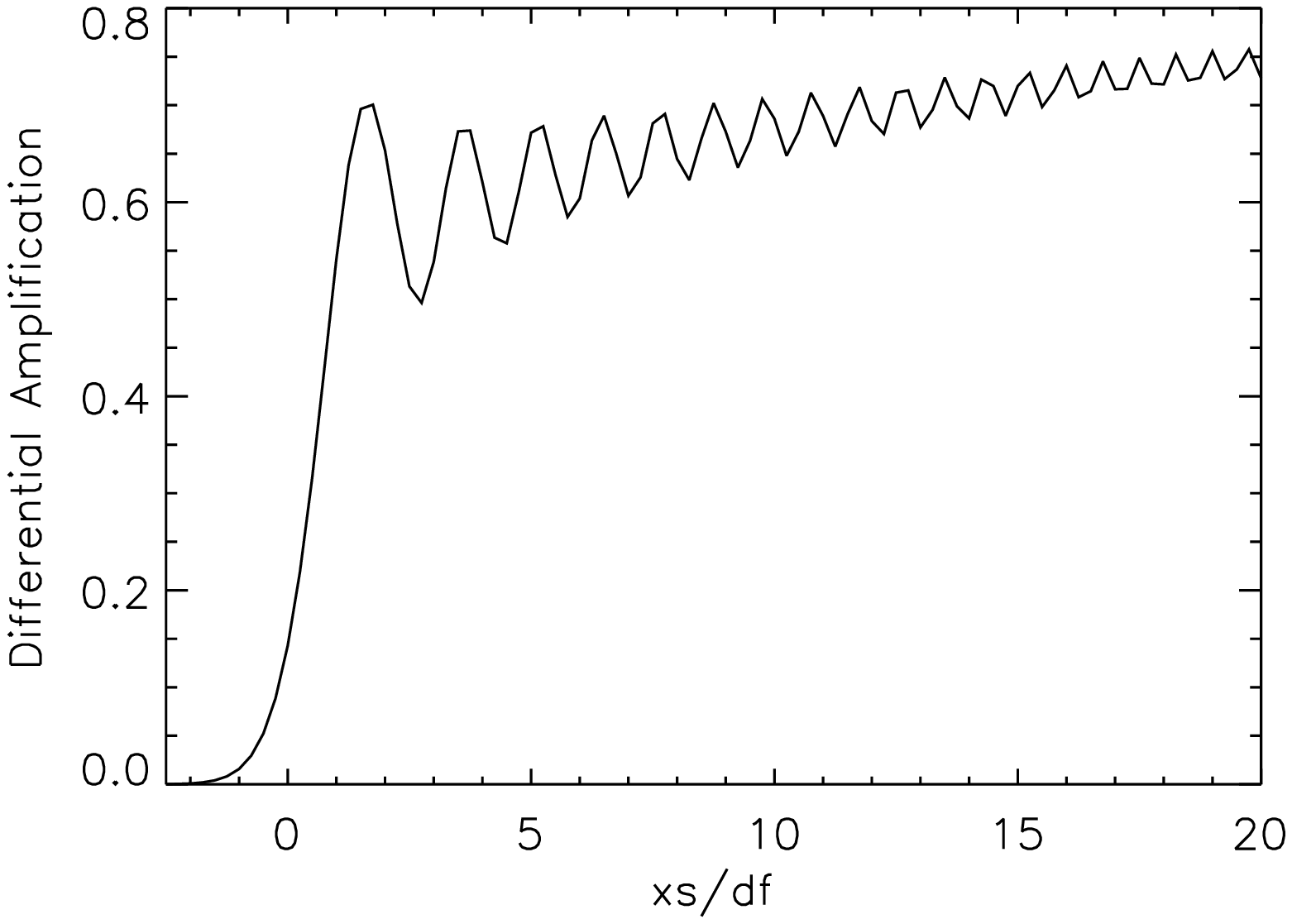} \caption{{\bfseries Magnification and differential magnification for a hypothetical white dwarf source in the 98-SMC-1 event.} $x_s$ is the distance the limb of the source has passed into the caustic. In the magnification plot this distance is scaled by the radius, in the differential plot it is scaled by the Airy diffraction length $d_f$. The corresponding scale expansion is $r_{source}/d_f$=240. The plots are sampled at $d_f$/4. The peak-to-peak amplitude of the diffraction signal in the differential plot is $\sim$ 40$\%$ of the unlensed source intensity.\label{fig10}}
\end{figure}


\clearpage
\begin{thebibliography}{}
\bibitem[Albrow et. al. (2001)]{alb01} Albrow, M.D., et. al. 2001, \apj, 549, 759
\bibitem[Alcock et. al. (2000)]{alc00} Alcock, C., et. al. 2000, \apj, 541, 270
\bibitem[Blandford $\&$ Narayan (1986)]{bla86} Blandford, R.D., and Narayan, R. 1986, \apj, 310, 568
\bibitem[Borra (1997)]{bor97} Borra, E.F. 1997, \mnras, 289, 660
\bibitem[Bozza (2000)]{boz00} Bozza, V. 2000, \aap, 374, 13
\bibitem[Carrol $\&$ Ostlie (1980)]{car80} Carrol, D., and Ostlie, Q. 1980, \textsl{An Introduction to Modern Astrophysics}, Addison Wesley
\bibitem[Deguchi $\&$ Watson (1986a)]{deg86} Deguchi, S., and Watson, W. 1986a, \prd, 34, 6, 1708
\bibitem[Deguchi $\&$ Watson (1986b)]{daw86} Deguchi, S., and Watson, W. 1986b, \apj, 307, 30
\bibitem[Goodman et. al. (1987)]{goo87} Goodman, J.J., Romani, R.W., Blandford, R.D., and Narayan, R. 1987, \mnras, 229, 73
\bibitem[Gould (1992)]{gou92} Gould, A. 1992, \apj, 386, L5
\bibitem[Gould (2001)]{gou01} Gould, A. 2001, \pasp, 113, 903
\bibitem[Jaroszy\'nski $\&$ Paczy\'nski (1995)]{jar95} Jaroszy\'nski, M., and Paczy\'nski, B. 1995, \apj, 455, 443
\bibitem[Kayser $\&$ Witt (1989)]{kay89} Kayser, R. and Witt, H.J. 1989, \aap, 221, 1
\bibitem[Kuhn at. al. (1998)]{kun98} Kuhn, J.R., Bush, R.I., Scheick, X., and Scherrer, P. 1998, \nat, 392, 155
\bibitem[Lennon et. al. (1996)]{len96} Lennon, D.J., Mao, S., Fuhrmann, K., and Gehren, T. 1996, \apj, 471, L23
\bibitem[Lites (1983)]{lit83} Lites, B.W. 1983, \solphys, 85, 193L
\bibitem[Ohanian (1982)]{oha82} Ohanian, H.C. 1982, \apj, 271, 551
\bibitem[Peterson $\&$ Falk (1991)]{pet91} Peterson, J.B. and Falk, T 1991, \apj, 374, L5
\bibitem[Petters, Levine, $\&$ Wambsganss (2001)]{pet01} Petters, A.O., Levine, H., and Wambsganss, J. 2001, \textsl{Singularity Theory and Gravitational Lensing}, Birkh$\ddot{a}$user
\bibitem[Refsdal (1964)]{ref64} Refsdal, S. 1964, \apj, 128, 295
\bibitem[Rhie $\&$ Bennett (2002)]{rhi02} Rhie, S.H., and Bennett, D. 2002, \apj, in press [astro-ph/9912050]
\bibitem[Rhie et. al. (1998)]{rhi98} Rhie, S.H., et. al. 1998, \apj, in press [astro-ph/9812252]
\bibitem[Rhie et. al. (1999)]{rhi99} Rhie, S.H., et. al. 1999, \apj, 522, 1037
\bibitem[Schneider, Elhers, $\&$ Falco (1992)]{sef92} Schneider, P., Elhers, J., and Falco, E.E. 1992, \textsl{Graviational Lenses}, Springer-Verlag
\bibitem[Stanek, Paczy\'nski, $\&$ Goodman (1993)]{sta93} Stanek, K.Z., Paczy\'nski, B., and Goodman, J. 1993, \apj, 413, L7
\bibitem[Udalski et. al. (1992)]{uda92} Udalski, A., Szyma\'nski, M., Kalu\'zny, J., Kubiak, M., and Mateo, M. 1992, \aca, 42, 253
\bibitem[Udalski et. al. (1997)]{uda97} Udalski, A., Kubiak, M., and Szyma\'nski, M. 1997, \aca, 47, 319
\bibitem[Ulmer $\&$ Goodman (1995)]{ulm95} Ulmer, A., and Goodman, J. 1995, \apj, 442, 67
\bibitem[Witt (1990)]{wit90} Witt, H.J. 1990, \aap, 236, 311
\bibitem[Witt, Kayser, $\&$ Refsdal (1993)]{wit93} Witt, H.J., Kayser, R., and Refsdal, S. 1993, \aap, 268, 501
\end{thebibliography}
\end{document}